\documentclass[aps,twocolumn,prl,showpacs,superscriptaddress]{revtex4}  
\usepackage{graphicx}   
\usepackage{dcolumn}    
\usepackage{bm}         
\usepackage{amssymb}    
\newcommand{\nc}{\newcommand}
\nc{\zev}{\lambda_0^{(V)}} \nc{\cO}{{\mathcal O}} \nc{\cP}{{\mathcal
P}} \nc{\dPdnu}{\frac{d\cP(\mu)}{d\mu}} \nc{\Av}[1]{\langle {#1}
\rangle} \nc{\nn}{\nonumber}

\begin{document}
\title{Drift instability and tunneling of lattice solitons}
\author{Y.~Sivan}
\affiliation{Department of Physics and Astronomy, Tel Aviv
University, Israel}

\author{G.~Fibich}
\affiliation{Department of Applied Mathematics, Tel Aviv University,
Israel}

\author{B. Ilan}
\affiliation{School of Natural Sciences, University of California,
Merced, California, USA}


\begin{abstract}
We derive an analytic formula for the lateral dynamics of solitons
in a general inhomogeneous nonlinear media, and show that it can be
valid over tens of diffraction lengths. In particular, we show that
solitons centered at a lattice maximum can be ``mathematically
unstable'' but ``physically stable''. We also derive an analytic
upper bound for the critical velocity for tunneling, which is valid
even when the standard Peierls-Nabarro potential approach fails.
\end{abstract}

\pacs{42.65 Jx, 42.65 Tg, 03.75 Lm} \maketitle

Solitons have been thoroughly studied in diverse fields of physics
such as nonlinear optics, BEC, plasma and water waves. By now, the
stability and dynamics of solitons in homogeneous media are well
understood. The possibility to manufacture transparent materials
with spatially varying, high contrast dielectric properties raises
new questions regarding stability and dynamics of solitons in
inhomogeneous media. In particular, while in a homogeneous medium
the solitons can freely move sideways, in an inhomogeneous medium
the loss of translation invariance affects the lateral movement of
solitons. The problem of lateral movement of lattice solitons is
interesting theoretically and important for applications such as all
optical switching and quantum information science. It was studied
analytically, numerically and experimentally for various media and
lattices, see
e.g.,~\cite{misc-mobility,Kivshar-PNP,PNP-1d-step,NLS_NL_MS,narrow_lattice_solitons,delta_pot_complete}.
However, each of these studies considered a specific nonlinearity,
lattice type and dimension.

In this Letter, we provide, apparently for the first time, a unified theory for the mobility of
lattice solitons which is valid for any nonlinearity, lattice type
and dimension. We show that soliton
mobility is intrinsically related to soliton stability, two key
properties that so far were studied separately. This relation
enables us to compute analytically the rate of drift of solitons
initially centered near a lattice maximum, and the restoring force
that the lattice exerts upon solitons initially centered near a
lattice minimum. In the latter case, our approach provides an upper
bound for the critical velocity for tunneling, which is valid even
when the standard Peierls-Nabarro potential approach cannot be
applied.

All solitons centered at a lattice maximum are ``{\em mathematically
unstable}'', as they drift towards the nearest lattice
minimum~\cite{Pelinovsky-04,NLS_NL_MS,delta_pot_complete,narrow_lattice_solitons}.
However, the ability to compute the magnitude of the drift rate
allows us to identify cases in which the drift rate is so small so
that the soliton is ``{\em physically stable}'', i.e., the drift
instability does not develop over the propagation distance of the
experiment. This observation explains why in some experiments
solitons centered at a lattice maximum were observed to be
stable~\cite{PNP-1d-step}. 

Consider the dimensionless nonlinear Schr\"odinger equation (NLS)
with a linear lattice
\begin{equation}
iA_z(z,{\bf x}) = - \nabla^2 A - F\left(|A|^2\right) A + V(N{\bf x})
A, \label{eq:NLS_inhom}
\end{equation}
where ${\bf x} = (x_1,\dots,x_d)$ and $V$ is a linear
lattice/potential with a characteristic length scale or period
$1/N$. This equation describes propagation in a media with Kerr
nonlinearity as well as quintic, cubic-quintic,
saturated/photorefractive nonlinearities etc. The variable $z$
denotes the propagation coordinate in nonlinear optics and time
coordinate in BEC. Eq.~(\ref{eq:NLS_inhom}) has soliton solutions $A
= e^{ i \mu z} u({\bf x})$, where $u$ is the solution of
\begin{equation}
\nabla^2 u({\bf x}) + F(|u|^2)u - V(N{\bf x}) u - \mu u = 0.
\label{eq:bs}
\end{equation}
It is well known that a necessary condition for stability is the
slope (VK) condition $\frac{d \mathcal{P}}{d \mu} > 0$ where
$\mathcal{P} = \int u^2 d{\bf x}$ is the soliton power. Violation of
the slope condition leads to a {\em width instability}, i.e., small
perturbations can lead to large changes of the soliton width, which
in some cases result in
collapse~\cite{NLS_NL_MS,narrow_lattice_solitons,delta_pot_complete}.

If the soliton is centered at a lattice maximum, it can become
unstable even if the slope condition is
satisfied~\cite{NLS_NL_MS,Pelinovsky-04,narrow_lattice_solitons,delta_pot_complete,PNP-1d-step}.
Indeed, there is a second condition for stability, the {\em spectral
condition}, which states that for $A = u e^{i \mu z}$ to be stable,
the number of negative eigenvalues of the operator $L^{(V)}_{+,\mu}
= -\nabla^2 + \mu + V({\bf x}) - F(u^2) - 2u^2F'$ should be at most
one more than the number of negative eigenvalues of the operator
$L^{(V)}_{-,\mu} = -\nabla^2 + \mu + V({\bf x}) -
F(u^2)$~\cite{spectral-cond}.

A simpler version of the spectral condition was derived
in~\cite{NLS_NL_MS} for the case $u>0$. Recall that in a homogeneous
medium, $L^{(V\equiv0)}_{+,\mu}$ has $d$ zero eigenvalues
$\lambda_{0,j}^{(V\equiv0)} = 0$ with corresponding eigenfunctions
$f_j^{(V\equiv 0)} = \frac{\partial Q}{\partial x_j}$, where $Q =
u^{(V \equiv 0)}$ is the solution of~(\ref{eq:bs}) with $V \equiv
0$. The potential $V$ breaks the translation symmetry of the medium.
As a result, $\lambda_{0,j}^{(V)}$ can split into $d$ different
values. The spectral condition is violated if and only if at least
one $\lambda_{0,j}^{(V)}$ attains a negative value~\cite{NLS_NL_MS}.

The spectral condition can be derived from the following linear
stability analysis~\cite{spectral-cond}. Let $A = e^{i\mu
z}\left(u({\bf x}) + h(z,{\bf x}) \right)$ where $h(z,{\bf x})$ is
a small perturbation. Since the instability due to violation of
the spectral condition originates only from the eigenfunctions
$f^{(V)}_j({\bf x})$ of $L_{+,\nu}^{(V)}$ which correspond to
negative $\lambda_{0,j}^{(V)}$, we can rewrite the perturbation
$h$ as
$$
h(z,{\bf x}) = c_1 e^{\Omega_j z}(f^{(V)}_j + ig^{(V)}_j) + c_{-1}
e^{-\Omega_j z}(f^{(V)}_j - ig^{(V)}_j),
$$ 
where $c_{\pm 1}$ are constants and 
$L^{(V)}_{+,\mu} f^{(V)}_j = - \Omega_j g^{(V)}_j, \quad \quad
L^{(V)}_{-,\mu} g^{(V)}_j = \Omega_j f^{(V)}_j$. 
Since $L^{(V)}_{+,\mu} f^{(V)}_j = \lambda_{0,j}^{(V)} f^{(V)}_j$
and ${L^{(V)}_{-,\mu}}$ is positive definite,
\begin{eqnarray}\label{eq:RQ-scaling}
\Omega_j^2 = - C_V \lambda_{0,j}^{(V)}, \quad C_V = \frac{\langle
f^{(V)}_j,f^{(V)}_j\rangle }{\langle {L^{(V)}_{-,\mu}}^{-1}
f^{(V)}_j,f^{(V)}_j\rangle} > 0.
\end{eqnarray}
Therefore, $\Omega_j$ is real (i.e., instability) when
$\lambda_{0,j}^{(V)} < 0$ and imaginary (i.e., stability) when
$\lambda_{0,j}^{(V)} > 0$~\cite{spectral-cond}.

The effect of a lattice on $\lambda_{0,j}^{(V)}$ was studied in
e.g.,~\cite{Pelinovsky-04,NLS_NL_MS,delta_pot_complete,narrow_lattice_solitons,Rapti_Kev_jones},
where it was shown that if the soliton is centered at a lattice
minimum (maximum), then $\lambda_{0,j}^{(V)}$ becomes positive
(negative), hence the spectral condition is satisfied (violated).

In~\cite{Pelinovsky-04,NLS_NL_MS,delta_pot_complete,narrow_lattice_solitons}
it was observed numerically that violation of the spectral condition
results in a drift instability, i.e., 
the center of mass (COM) of the beam in the $x_j$ coordinate,
defined as $\overline{x_j}(z) = \int x_j|A|^2 / \cP$, drifts away
from its initial location $\overline{x_j}(0)$ near the lattice
maximum.
{\em So far, however, the relation between the spectral condition and the
drift instability has not been established analytically}. To do that,
we note that since $f^{(V)}_j$ and $g^{(V)}_j$ are
odd~\cite{NLS_NL_MS}, 
\begin{eqnarray}\label{eq:com_general}
\overline{x_j}(z) &=& \frac{1}{\cP}\Av{x_j, \left|u({\bf x};\mu) +
h(z,{\bf x}) \right|^2 } \\
&=& B \left(c_1 e^{\Omega_j z} + c_{-1} e^{- \Omega_j z} \right),
\nn
\end{eqnarray}
where $B = 2 \Av{x_j, u f^{(V)}_j} / \cP$ is constant. 
Thus, 
\begin{equation}\label{eq:com_dyn}
\ddot{\overline{x_j}}(z) = \Omega_j^2 \overline{x_j}(z).
\end{equation}
Relation~(\ref{eq:com_dyn}) shows that {\em a failure to satisfy the
spectral condition ($\lambda_{0,j}^{(V)} < 0$) leads to a drift
instability. Moreover, the {\em magnitude of $\Omega_j = |C_V
\lambda_{0,j}^{(V)}|^{\frac{1}{2}}$} determines the drift rate away
from} the lattice maximum in the $x_j$~direction.

A simpler expression for $C_V$ can be obtained for a weak lattice
($V \ll \mu$) and a power nonlinearity $F = |A|^{p-1}$. In this
case, $f^{(V)}_j \cong f^{(V \equiv 0)}_j = \frac{\partial
Q}{\partial x_j}$ and $L^{(V)}_{-,\mu} \cong L^{(V \equiv
0)}_{-,\mu} = \mu L^{(V \equiv 0)}_{-,1}$. Thus, $C_V \cong \mu
\langle\frac{\partial Q}{\partial x_j},\frac{\partial Q}{\partial
x_j}\rangle / \langle L_{-,\mu}^{-1} \frac{\partial Q}{\partial
x_j},\frac{\partial Q}{\partial x_j}\rangle\big|_{\mu=1}$. From the
Pohozaev identities it follows that $\Av{\frac{\partial Q}{\partial
x_j},\frac{\partial Q}{\partial x_j}} = \frac{1}{4\delta} \Av{Q,Q}$
where $\delta = \frac{(2-d) p + 2 + d}{4(p-1)}$. In addition, if we
multiply $L_{-,\mu} w = \frac{\partial Q}{\partial x_j}$ by $x_j Q$
and integrate in parts we get $\Av{L_{-,\mu}^{-1} \frac{\partial
Q}{\partial x_j},\frac{\partial Q}{\partial x_j}} = \frac{1}{4}
\Av{Q,Q}$. Substituting in~(\ref{eq:RQ-scaling}) gives
\begin{equation}\label{eq:RQ-weak}
\Omega_j^2 \cong - \frac{\mu}{\delta} \lambda_{0,j}^{(V)}. 
\end{equation}
Hence, for a weak lattice, the dependence of the drift rate
$\Omega_j$ on the lattice period $N$ is only through its effect on
$\zev$. The approximation~(\ref{eq:RQ-weak}) can be generalized for
different nonlinearities and to lattices which are not weak. For
example, in the case of narrow lattice, $\Omega_j^2$ is given
by~(\ref{eq:RQ-weak}) with $\mu$ replaced by $\mu + V({\bf x}_0)$
where ${\bf x}_0$ is the location of the soliton
peak~\cite{narrow_lattice_solitons}.

We solve Eq.~(\ref{eq:NLS_inhom}) numerically for $F = |A|^2$, $d =
1$ and
\begin{equation}\label{eq:cos_lattice}
V(x) = V_0 \cos(2 \pi N x),
\end{equation}
with the initial condition $A(0,x) = u(x - \delta)$, where $u(x)$ is
the solution of~(\ref{eq:bs}) centered at $x=0$. Therefore,
$\overline{x}(z=0) = \delta$ (since $d=1$, we can suppress the index
$j$). For a small shift $\delta \ll 1$, we can rewrite $A(0,x) =
u(x) + h(0,x)$ where $h(0,x) = - \delta \frac{du}{dx} +
\cO(\delta^2)$. Since $h(0,x) = (c_1 + c_{-1})f^{(V)} + i (c_1 -
c_{-1}) g^{(V)}$, then $c_1 - c_{-1} = 0$ and
by~(\ref{eq:com_general}),
\begin{eqnarray}\label{eq:cosh}
\overline{x}(z) &=& \delta \cosh{\Omega z}.
\end{eqnarray}

\begin{figure}[ht!]
\includegraphics[width=9cm]{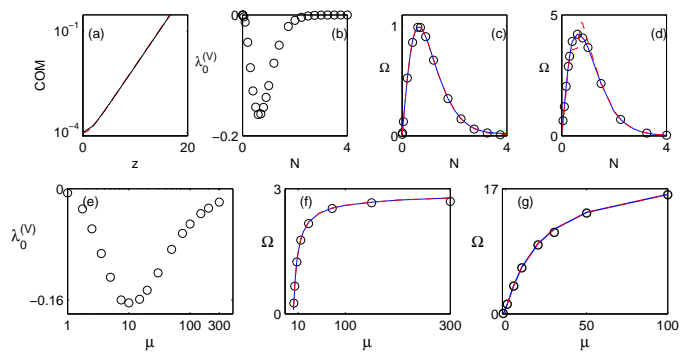} \caption{
(Color online) (a) The dynamics of the COM for $V_0 = 0.1$, $\delta
= 10^{-4}$ and the lattice~(\ref{eq:cos_lattice}) for $N=0.2$
(dashed red line), $N=1.437$ (dotted blue line), and the analytical
prediction~(\ref{eq:cosh}) with $\Omega \sim 0.52$ (black solid
line). The 3 lines are indistinguishable. (b) $\zev$ as a function
of $N$. (c) Drift rate $\Omega$ as a function of~$N$. The analytical
prediction~(\ref{eq:RQ-scaling}) (solid blue line) and its
approximation~(\ref{eq:RQ-weak}) (dashed red line) are nearly
indistinguishable. (d) Same as (c) for $V_0 = 2$. (e) $\zev$ as a
function of $\mu$. (f) Same as (c), but as a function of $\mu$. (g)
Same as (f) for $d=2$ and the lattice~(\ref{eq:coscos}) with $V_0 =
5$. } \label{fig:drift_N_mu}
\end{figure}

In our simulations we observe that that the COM evolves according
to~(\ref{eq:cosh}), see Fig.~\ref{fig:drift_N_mu}(a), and therefore,
calculate the drift rate numerically by finding the best fitting
$\Omega$. We fix $V_0 = 0.1$ and $\mu = 4.5$ and vary~$N$. As
expected for a soliton centered at a lattice
maximum~\cite{narrow_lattice_solitons,NLS_NL_MS}, $\zev < 0$ for all
values of~$N$, and $\zev$ vanishes in the limits $N \to 0$ (narrow
solitons) and $N \to \infty$ (wide solitons), see
Fig.~\ref{fig:drift_N_mu}(b). In Fig.~\ref{fig:drift_N_mu}(c) we
confirm that the numerically computed drift rate $\Omega$ is in
excellent agreement with Eq.~(\ref{eq:RQ-scaling}) and also with the
approximation~(\ref{eq:RQ-weak}). Accordingly, each value of $\zev$
is attained at two different values of $N$ for which the drift rates
are nearly identical. Indeed, in Fig.~\ref{fig:drift_N_mu}(a) we see
that the drift rate of the COM when $N = 0.2$ and $N = 1.437$, both
for which $\zev \cong - 0.0454$, is the same over more than 3 orders
of magnitude and 40 diffraction lengths. In
Fig.~\ref{fig:drift_N_mu}(d) we repeat these simulations with a
stronger lattice ($V_0 = 2$). In this case, the numerically computed
drift rate is in excellent agreement with the one predicted by
Eq.~(\ref{eq:RQ-scaling}). The approximation~(\ref{eq:RQ-weak}) is
very accurate only for narrow ($N \ll 1$) and wide ($N \gg 1$)
solitons. 
Indeed, although the lattice oscillations are not small, for narrow
and wide solitons, the effect of a mean-zero lattice is weak, hence
the deviation of $f^{(V)}_j$ from $\frac{\partial Q}{\partial x_j}$
is small~\cite{NLS_NL_MS,narrow_lattice_solitons}. Although for
solitons of $N = \cO(1)$ width the deviation of $f^{(V)}_j$ from
$\frac{\partial Q}{\partial x}$ is not small, the
approximation~(\ref{eq:RQ-weak}) is, at most, $10\%$ inaccurate.

In Fig.~\ref{fig:drift_N_mu}(e) we fix $N = 1$ and $V_0 = 0.1$ and
vary $\mu$. As in Fig.~\ref{fig:drift_N_mu}(b), since the soliton is
centered at a lattice maximum, $\zev < 0$ for all values of $\mu$,
and $\zev$ vanishes in the two limits $\mu \to 0$ (wide solitons)
and $\mu \to \infty$ (narrow solitons). In
Fig.~\ref{fig:drift_N_mu}(f) we see that $\Omega$ is monotonically
increasing in $\mu$, and that the numerically calculated drift rate
is in excellent agreement with the analytical
prediction~(\ref{eq:RQ-scaling}) and also with its
approximation~(\ref{eq:RQ-weak}).

We emphasize that despite the similarity of the dependence of $\zev$
on $N$ and $\mu$ (see Fig.~\ref{fig:drift_N_mu}(b) and
Fig.~\ref{fig:drift_N_mu}(e)), the dependence of $\Omega$ on $N$ and
$\mu$ is completely different in the narrow-beam limit. Indeed, for
narrow beams $\lambda_0^{(V)} \cong 4 \delta \frac{N^2}{\mu}
\frac{d^2 V}{dx^2}\big|_{x=0}$~\cite{narrow_lattice_solitons} so
that by~(\ref{eq:RQ-weak}), $\Omega^2 \approx \Omega^2_{narrow} = -
4 N^2 \frac{d^2 V}{dx^2}\big|_{x=0}$. Hence, $\Omega$ vanishes for a
fixed $\mu$ and $N \to 0$ (Fig.~\ref{fig:drift_N_mu}(c)) but
approaches $\Omega_{narrow} \cong 2.8$ for a fixed $N$ and $\mu \to
\infty$ [Fig.~\ref{fig:drift_N_mu}(f)].

In order to show that our results are also valid in higher
dimensions, we solve Eq.~(\ref{eq:NLS_inhom}) in a $d=2$ setting
with
\begin{equation}\label{eq:coscos}
V(x,y) = \frac{V_0}{2} \left(cos^2(2 \pi x) + cos^2(2 \pi y)\right),
\end{equation}
with $V_0 = 5$, and find the numerical drift rate to be in excellent
agreement with Eq.~(\ref{eq:RQ-scaling}), see
Fig.~\ref{fig:drift_N_mu}(g). Remarkably, although the lattice is
strong, the numerical drift rate is also in excellent agreement with
the approximation~(\ref{eq:RQ-weak}) in which $\mu$ is shifted by
$V_0/2$, the mean of~$V$.

The analytical relation~(\ref{eq:cosh}), together
with~(\ref{eq:RQ-scaling}) or~(\ref{eq:RQ-weak}), enable us to
estimate the distance at which a soliton initially centered near a
lattice maximum will deviate significantly from its initial
location. In particular, if the initial shift~$\delta$ and/or drift
rate $\Omega$ are sufficiently small, then this ``mathematically
unstable'' soliton can remain ``near'' its initial location over the
propagation distance of the experiment, i.e., be ``physically
stable''. This observation can explain the experimental results
of~\cite{PNP-1d-step}, where solitons centered at a lattice maximum
did not drift over $\approx 18$ diffraction lengths.

{\em The relation between the sign and magnitude of the perturbed
near-zero eigenvalues $\{\lambda^{(V)}_{0,j}\}_{j=1}^d$ and the
drift instability appears to be universal}. Indeed, we now show that
it also occurs in numerical calculation of soliton profiles using
Petviashvili's iterations method (PIM), which is nowadays frequently
used in optics and BEC~\cite{Fourier-iterations}.
In~\cite{Fourier-iterations-proof} it was proved that PIM converges
only if $L^{(V)}_{+,\mu}$ has at most one negative eigenvalue, which
is a spectral condition similar to the one for the stability of NLS
solitons. Accordingly, PIM is not expected to converge for solitons
centered at lattice maxima.

\begin{figure}[ht!]
\centerline{\includegraphics[width=5.5cm]{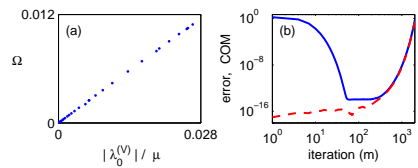}}
\caption{ (Color online) (a) Drift rate $\Omega$ as a function of
$\zev / \mu$ for the solution of~(\ref{eq:bs}) using PIM. (b)
Maximal error (solid line) and COM (dashed red line) in the $m$th
iteration. } \label{fig:SL_SR}
\end{figure}

We solve Eq.~(\ref{eq:bs}) with $F = |u|^2$ using PIM with the
initial guess $u^{(0)} = u(x - \delta)$, where $u(x)$ is the
solution of~(\ref{eq:bs}) centered at a maximum of the
lattice~(\ref{eq:cos_lattice}) with $V_0 = 0.1$. Similarly to the
dynamics of NLS solitons centered slightly off a lattice maximum,
the COM of $u^{(m)}$ evolves according to $\overline{x}(m) \sim
\delta e^{\Omega m}$ (data not shown), where $u^{(m)}$ is the
solution in the $m$th iteration. Thus, we conclude that when the
spectral condition for PIM is violated, the method does not converge
because the iterative solution drifts away from the lattice maximum.
In that sense, the analogy between the dynamics (in $z$) of NLS
solitons and of $u^{(m)}$ (in $m$) is further demonstrated, since in
both cases, violation of the spectral condition leads to a drift
instability. We also compute the exponential drift rate~$\Omega$
numerically for various combinations of $N$ and $\mu$ and observe
that $\Omega^2 \cong - D_V \left(\zev(N,\mu)/\mu\right)^2$
where the constant $D_V$ depends on $V_0$ but is independent of
$\mu$ and $N$, see Fig.~\ref{fig:SL_SR}(a). Interestingly, the
scaling of $\Omega$ in $\zev$ and in $\mu$ is different
from~(\ref{eq:RQ-weak}), yet in both cases $\Omega$ depends on $N$
only through~$\zev$.

Although in~\cite{Fourier-iterations-proof} it was proved that the
iterations should diverge for solitons centered at a lattice
maximum, in several studies these iterations did
``converge''~\cite{NLS_NL_MS,narrow_lattice_solitons,Ablowitz_Ilan_irreg_lattices}.
To explain this apparent inconsistency, in Fig.~\ref{fig:SL_SR}(b)
we plot $max_x|u^{(m)} - u|$ as a function of $m$ for $N = 0.2$,
$\mu = 2$ and the lattice~(\ref{eq:cos_lattice}) with $V_0 = 0.1$
and $u^{(0)} = e^{-x^2}$, and observe that the iterations converge
(i.e., $max_x|u^{(m)} - u| < 10^{-13}$) after $\approx 40$
iterations. However, if we continue the iterations, a significant
drift of the COM occurs around $m \approx 2000$. To understand this
``post-convergence'' drift, we note that in this example, $\zev
\cong - 0.085$ and $\Omega \cong 0.018$. Since the seed of the drift
is roundoff error, then $\overline{x}(m=0) = \cO(10^{-16})$. Indeed,
$10^{-16}e^{0.018\cdot 2000} = \cO(1)$. This example of a numerical
iterative solution which theoretically should diverge yet in
practice converges is thus analogous to the mathematically unstable
yet physically stable NLS solitons discussed earlier.

We now consider solitons centered near a lattice minimum. Since
$\lambda_{0,j}^{(V)} > 0$, the spectral condition is satisfied.
Hence, these solitons are stable under small lateral perturbations.
Indeed, relation~(\ref{eq:com_dyn}) shows that small lateral
perturbations would lead to oscillations around the lattice maximum,
while relation~(\ref{eq:RQ-scaling}) shows that the magnitude of
$\lambda_{0,j}^{(V)}$ determines the strength of the restoring
force. For example, consider a soliton centered at a lattice minimum
which is launched at an angle $\theta_j$ between the $x_j$ and $z$
axes. Such an angle corresponds to an initial transverse velocity of
$v_{0,j} = \dot{\overline{x_j}}(z=0) = \tan \theta_j$. By
(\ref{eq:RQ-scaling})-(\ref{eq:com_general}), the COM evolves
according to
\begin{equation}\label{eq:sin}
\overline{x_j}(z) = v_{0,j} \sin\left(|\Omega_j| z\right) /
|\Omega_j|.
\end{equation}
Thus, as $\lambda_{0,j}^{(V)}$, hence $|\Omega_j|$, increase, the
maximal deviation of the COM from the lattice minimum becomes
smaller, implying stronger lateral stability.

\begin{figure}[ht!]
\centerline{\includegraphics[width=9cm]{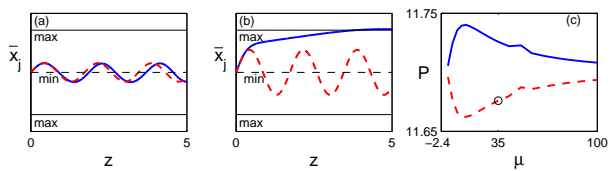}}
\caption{ (Color online) Dynamics of $\bar{x}_1(=\bar{x}_2)$ (solid)
and theoretical prediction~(\ref{eq:sin}) (dashes)  in a bulk medium
with $F = |A|^2$, the lattice~(\ref{eq:coscos}) with $V_0 = 0.5$,
$\mu = 35$ and (a) ${\bf v}_0 = (0.2,0.2)$, (b) ${\bf v}_0 =
(0.485,0.485)$. (c) Power of solitons centered at a lattice maximum
(solid) and minimum (dashes). } \label{fig:Tunnel_Kerr}
\end{figure}

Eq.~(\ref{eq:sin}) gives an accurate description of the dynamics for
small velocities. However, for non-small velocities, as the soliton
propagates sideways, the attraction towards the lattice minimum
decreases, an effect which is not captured by Eq.~(\ref{eq:sin}). To
see that, we solve Eq.~(\ref{eq:NLS_inhom}) with $d=2$, $F = |A|^2$
and the lattice~(\ref{eq:coscos}) with $V_0 = 0.5$. The initial
condition is $A(0,x,y) = u(x,y)e^{i (v_0 x + v_0 y) / 2}$, i.e., a
soliton centered at a lattice minimum ${\bf x}_{min} =(0,0)$ with
initial velocity in the direction of the nearest lattice maximum at
${\bf x}_{max} = (0.25,0.25)$. Indeed, for small initial velocities,
the agreement between the dynamics and Eq.~(\ref{eq:sin}) is
excellent, see Fig.~\ref{fig:Tunnel_Kerr}(a). For higher velocities,
the COM initially evolves according to Eq.~(\ref{eq:sin}) but
deviates from it as it approaches the lattice maximum, see
Fig.~\ref{fig:Tunnel_Kerr}(b). For a sufficiently large initial
velocity, the soliton can ``tunnel'' beyond the nearest lattice
maximum. The critical velocity for tunneling ${\bf v}_0^{cr}$ is the
one for which the transverse velocity $\dot{\bar{{\bf x}}}(z)$
vanishes at the lattice maximum. The upper limit $|{\bf v}_0^{cr}|
\le v_{th}^{cr} = \sqrt{\sum_{j=1}^d |\Omega_j|^2 ({\bf x}_{max} -
{\bf x}_{min})_j^2}$ can be derived from Eq.~(\ref{eq:sin}). In the
case of Fig.~\ref{fig:Tunnel_Kerr}(b), this bound gives $|{\bf
v}_0^{cr}| \le 1.22$, an over-estimate of~$\approx 45\%$ over $|{\bf
v}_0^{cr}| \approx 0.485 \sqrt{2}$ .

\begin{figure}[ht!]
\centerline{\includegraphics[width=9cm]{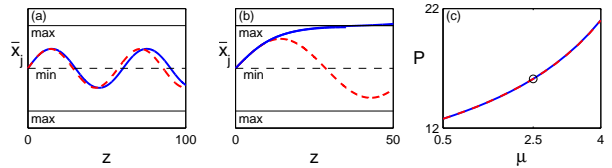}} \caption{
(Color online) Same as Fig.~\ref{fig:Tunnel_Kerr} for $F = |A|^2 -
0.02|A|^4$, $\mu = 2.5$, $V_0 = 1$ and (a) ${\bf v}_0 =
(0.0125,0.0125)$, (b)~${\bf v}_0 = (0.01905,0.01905)$. }
\label{fig:Tunnel_CQ}
\end{figure}

The standard formula for ${\bf v}_0^{cr}$, 
based on the Peierls-Nabarro potential (PNP)
approach~\cite{Kivshar-PNP}, is $|{\bf v}_0^{cr}| = \sqrt{4
\Delta\mathcal{H} / \mathcal{P}}$ where $\Delta \mathcal{H}$ is the
difference in the Hamiltonians of equal-power solitons centered at a
lattice minimum and maximum, respectively.
In order to apply the PNP approach, the power of the soliton
centered at a lattice maximum should be equal to that of a soliton
centered at a lattice minimum. For a two-dimensional Kerr medium ($F
= |A|^2$), however, such ``soliton pairs'' do not exist, since the
power of all solitons centered at a lattice maximum is below that of
all solitons centered at a lattice minimum, see
Fig.~\ref{fig:Tunnel_Kerr}(c). Therefore, one cannot use the PNP
approach, and the upper bound $v_{th}^{cr}$ provides the only
analytic estimate of $|{\bf
v}_0^{cr}|$. 

Finally, we solve Eq.~(\ref{eq:NLS_inhom}) for a cubic-quintic
nonlinearity  and the lattice~(\ref{eq:coscos}). As in the Kerr
case, for small initial velocities, the agreement between the
numerics and Eq.~(\ref{eq:sin}) is excellent over many diffraction
lengths [Fig.~\ref{fig:Tunnel_CQ}(a)], while for higher velocities
the COM deviates from Eq.~(\ref{eq:sin}) as the soliton approaches
the lattice maximum [Fig.~\ref{fig:Tunnel_CQ}(b)]. In this case the
PNP approach is applicable [see Fig.~\ref{fig:Tunnel_CQ}(c)] and
yields $|{\bf v}_0^{cr}| \simeq 0.027$. The value of the critical
velocity obtained numerically is within $1\%$ of the PNP prediction
[see Fig.~\ref{fig:Tunnel_CQ}(b)]. To the best of our knowledge,
this is the first demonstration of quantitative agreement of the PNP
approach with numerical results for $d=2$. The research of G.F. and
Y.S. was partially supported by BSF grant no. 2006-262.




\end{document}